# Nanoscale spin manipulation with pulsed magnetic gradient fields from a hard disc drive writer

S. Bodenstedt[1], I. Jakobi[1,*], J. Michl[1], I. Gerhardt[1,2], P. Neumann[1] and J.Wrachtrup[1]

[1]3. Physikalisches Institut, Universität Stuttgart and Institute for Integrated Quantum Science and Technology IQ[ST], Pfaffenwaldring 57, D-70569 Stuttgart, Germany

[2]Max Planck Institute for Solid State Research, Heisenbergstraße 1, D-70569 Stuttgart, Germany

**ABSTRACT:** The individual and coherent control of solid-state based electron spins is important covering fields from quantum information processing and quantum metrology to material research and medical imaging. Especially for the control of individual spins in nanoscale networks, the generation of strong, fast and localized magnetic fields is crucial. Highly-engineered devices that demonstrate most of the desired features are found in nanometer size magnetic writers of hard disk drives (HDD). Currently, however, their nanoscale operation, in particular, comes at the cost of excessive magnetic noise. Here, we present HDD writers as a tool for the efficient manipulation of single as well as multiple spins. We show that their tunable gradients of up to 100 µT/nm can be used to spectrally address individual spins on the nanoscale. Their GHz Bandwidth allows to switch control fields within nanoseconds, faster than characteristic timescales such as Rabi and Larmor periods, spin-spin couplings or optical transitions, thus extending the set of feasible spin manipulations. We used the fields to drive spin transitions through non-adiabatic fast passages or enable the optical readout of spin states in strong misaligned fields. Building on these techniques, we further apply the large magnetic field gradients for microwave selective addressing of single spins and show its use for the nanoscale optical colocalization of two emitters.

Networks of interacting *nitrogen-vacancy defect* (NV) spins in diamond have a wealth of potential applications[1], ranging from quantum information processing[2] to sensing arrays[3–5]. Spins in a coherently coupled network may exchange quantum information and form a quantum register. To fully control a network containing many NVs, the selective manipulation of individual spins has to be possible. However, when multiple NVs are located within the same diffraction limited spot, i.e. $d < 200$ nm, they cannot be easily controlled individually if they have parallel axis orientations. One needs to resort to super-resolution techniques like STED to individually address the defects which often have a negative impact on neighboring spins[6]. Magnetic field sources exhibiting strong gradient and a high bandwidth, i.e. fast switching times, offer an adequate solution. Gradient fields can be used to encode the location of NVs onto its spin properties such as its Larmor precession[7]. Hence, they separate these properties of otherwise indistinguishable spins and can make them individually addressable. Coherent dipolar coupling among neighboring spins requires distances below some tens of nanometers[8]. Detuning such spins necessitates strong gradients[7] on the order of 10-100 µT/nm. Fast switching times are required as the gradient field might interfere with the dynamics of the network, so that it can be exclusively applied for local manipulations of single spins. At the same time, fast changing control fields can also affect the spin dynamics, e.g. through non-adiabatic fast passages which have to be controlled or avoided. As a drawback, sources able to generate strong fields at a high bandwidth may also introduce strong magnetic noise. This effect needs to be taken into account, as it further limits the distance between coherently interacting spins.

We find the desired features for a gradient field source readily available in magnetic writers of commercial hard disk drives[9] (HDD). Modern HDDs can produce magnetic fields on the order of 1 T with gradients of up to 10 mT/nm and pulse them well within a nanosecond[10]. In order to flip the magnetization, i.e. write a bit, HDDs use a microscale lithographed electromagnet, the writer, as shown in Figure 1a. The demand for higher storage capacities and faster operation has prompted the miniaturization of the device and decades worth of engineering have developed HDD writers to a sophisticated tool.

HDD writers consist of a nickel-iron alloy core with a high permeability ($\mu > 10,000$) that is shaped in three poles perpendicular to the air-bearing surface (ABS) of the head and a connecting yoke on top. The central pole, the write pole, is tapered to a 100 nm wide tip on the ABS while the two outer (return) poles end in broad shielding brackets close to the write pole. The front return pole closes on the write pole with a small gap of about 20 nm, while the back return-pole ends in a few micrometer distance. A pair of coils is photolithographically designed around the yoke such that the write pole is jointly magnetized in one direction when a current of up to 30 mA is applied. Once magnetized, the magnetic field is emitted perpendicular to the ABS from the write pole and curls around to the nearby front return pole. This geometry creates a strong magnetic field gradient where only a 20 nm sized area is strong enough to flip the magnetization of the recording medium underneath the ABS. In order to write data quickly the writer is designed to have a GHz bandwidth. It was previously reported that NV spins are suitable nanoscale magnetometers for the characterization of such fields[10]. Here we reverse the situation and show their versatility in the control of single as well as multiple NV spins and demonstrate their potential use for the selective addressing of NV spins.

The NV is a point defect where one carbon atom is missing from the diamond lattice and a nitrogen atom substitutes a neighboring site (see Figure 1b). In its negative charge state the NV has triplet ($S = 1$) ground ($^3A$) and excited ($^3E$) states and additional metastable singlet ($S = 0$) states.

The Hamiltonian $\hat{H}$ describing the spin of the triplet ground state is in its simplest form[11]

$$\hat{H} = D\hat{S}_z^2 + \frac{\gamma}{2\pi}\left(B_x\hat{S}_x + B_y\hat{S}_y + B_z\hat{S}_z\right). \quad (1)$$

It exhibits a *zero-field splitting* (ZFS) of $D = 2.87$ GHz and an isotropic Zeeman interaction with a magnetic field $\vec{B}$ and a gyromagnetic ratio of $\gamma/2\pi = 28.02$ GHz/T. Figure 1c shows the spin energy levels for axial $B_z$ and radial $B_\perp = \left(B_x^2 + B_y^2\right)^{1/2}$ fields. At moderate magnetic fields $|\vec{B}| \ll 2\pi D/\gamma \approx 100$ mT the ZFS is the dominant term and defines the quantization axis along the symmetry axis $z$ of the defect. The contribution of axial fields $B_z\hat{S}_z$ leaves the quantization axis unchanged and only shifts the eigenenergies resulting in a linear Zeeman effect that splits the magnetic states $m_S = \pm 1$ by a factor $\gamma/\pi \cdot B_z$. Radial contributions $B_x\hat{S}_x$, $B_y\hat{S}_y$, on the other hand, mainly mix magnetic states $m_S =$

±1 and result in quadratic Zeeman interaction. With the knowledge of two transition frequencies two components of the magnetic field $B_z$ and $B_\perp$ and consequently the polar angle $\tan\theta = B_\perp/B_z$ can be drived while the azimuthal angle $\tan\phi = B_y/B_x$ remains ambiguous[12–15].

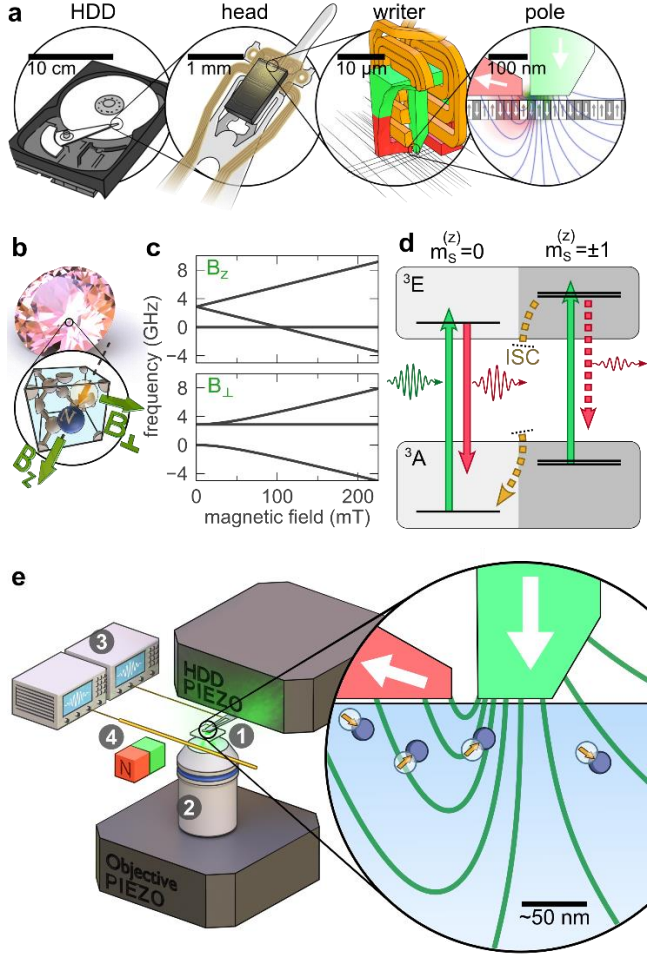

**Figure 1.** Hard disk drives (HDD) and nitrogen vacancy (NV) defects in diamond. **(a)** Schematic overview of a hard disk write head. In a HDD a head incorporating a writer is suspended over the disk. The writer is a microstructured electromagnet with a central pole emitting a field able to flip the magnetization of a 20 nm wide sector on the recording medium. **(b)** Diamond unit cell with NV defect showing the orientation of magnetic fields with respect to the defect symmetry axis. **(c)** Spin energy levels in the NV⁻ triplet ground state ³A for axial (top) and radial (bottom) magnetic fields. A zero-field-splitting of 2.87 GHz acting along the symmetry axis causes an anisotropic interaction with magnetic fields. **(d)** Electronic level scheme with optical cycle (green, red arrows) between triplet states ³A and ³E. Intersystem-crossing (ISC, yellow) allows a non-radiative decay to the singlet states (not depicted) and back to ³A resulting in a fluorescence contrast between spin states and spin polarization. **(e)** Experimental setup. A HDD head is mounted on a piezo stage and rests in flat contact on a diamond membrane (1). A confocal microscope (2) has access from below to monitor NV defects close to the writer (inset). The writer is controlled by a signal generator (3). A wire antenna and a permanent magnet are used for additional control fields (4).

Figure 1d depicts the electronic level scheme of the NV. A spin-conserving optical transition can be driven to excite the NV electron system into the ³E state. The spin state can be read out via the fluorescence intensity of the defect. Here the $m_S^{(z)} = 0$ state in the optically excited state mostly decays through fluorescence while the $m_S^{(z)} = \pm 1$ states have a significant rate of non-radiatively decaying through *intersystem crossing* (ISC)[16]. This leads to a fluorescence contrast between these spin states and allows for *optically detected magnetic resonance* (ODMR). Typically spin states are manipulated by resonant microwave radiation[17]. However, in radial fields the spin basis is mixed and all states have significant ISC rates. Therefore, ODMR only yields a sufficient fluorescence contrast under the constraint that radial fields are smaller than ~20 mT[18,19].

A schematic of the experimental setup is depicted in Figure 1e. We connect a generic perpendicular magnetic recording head to a frequency generator. In order to have the magnetic field of the writer interact with NV spins we place it with the ABS in flat contact to the surface of a diamond substrate. As the field strength of the writer decays strongly over distance we use NV defects 5-20 nm below the diamond surface. Using diamond membranes of 30-50 μm thickness, we have optical access from the back surface of the diamond. The membranes are mounted on an objective-scanning confocal microscope (60x/1.35 NA oil objective) to excite the electronic transitions and to collect the *photoluminescence* (PL). The HDD head together with its assembly is mounted on a separate piezo stage. The assembly's flexibility helps to stabilize the ABS's position with respect to the diamond surface as the head will mostly be kept in place by static friction. In order to actually move the head, the tension in the assembly has to overcome the static friction. At this point the head will start to slide over the diamond surface and can follow steps with a lateral positioning accuracy on the order of 10 nm.

In order to control individual spins on the nanoscale using the HDD's gradient field, the writer needs to be placed close to the NVs. However, while its ferromagnetic core enables the HDD's field amplitude and gradient, it comes at the cost of large magnetic noise even when the write current is off. Many applications of the NV rely on spin polarization or coherent phase evolution where the relaxation times $T_1$ and $T_2$ are the limiting factors, respectively. Hence, the impact of this noise source on the spin relaxation times needs to be characterized for such applications. To probe the noise spectrum in different frequency domains we employ relaxometry[20–24]. By measuring the longitudinal relaxation time $T_1 = 1/\Gamma_1$ we gain insight into the noise spectral density at the zero field Larmor frequency of about 3 GHz of noise components perpendicular to $z$. Measurements of the transversal relaxation time $T_2 = 1/\Gamma_2$ probe noise components at lower frequencies[25] parallel to $z$. The corresponding total relaxation rates $\Gamma_{\text{tot}} = \Gamma_{\text{int}} + \Gamma_{\text{ext}}$ can be separated into an intrinsic contribution $\Gamma_{\text{int}}$ caused by a variety of sources within the diamond and an external part $\Gamma_{\text{ext}}$ caused by the writer. For magnetic noise the external rate can be written as[25]

$$\Gamma_{\text{ext}} = \gamma^2 \cdot \langle B^2 \rangle \cdot \int S(\nu) \cdot F_i(\nu) \, d\nu, \qquad (2)$$

where $\gamma$ is the gyromagnetic ratio, $\sqrt{\langle B^2 \rangle}$ the effective magnetic field at the position of the NV, $S(\nu)$ the normalized *power spectral density* (PSD) depending on the frequency $\nu$. The filter function $F_i(\nu)$ describes the spectral sensitivity depending on the measurement sequence.

For the scan in Figure 2a the microscope objective is aligned to a single NV while for different writer positions the spin relaxation is measured to determine $T_1$. During the measurement no electric current is applied to the writer. With an increasing distance to the write and return poles the longitudinal spin relaxation time reaches several hundred microseconds (bottom part of Figure 2a and line plot in Figure 2b). Below the write and return pole the magnetic field noise reduces $T_1$ to values around 100 μs (upper part). Close to the expected position of the write pole (center in Figure 2a and

b) the longitudinal relaxation times are further reduced to only tens of microseconds. These two sets of data with and without influence of the writer allow us to estimate the magnetic noise.

The filter function $F_i(\nu)$ of a $T_1$ relaxometry measurement is only significant around the NV's ESR frequency around 3 GHz[25]. Assuming a constant PSD over this sensitivity range the spectral magnetic field noise $\langle B^2 \rangle \cdot S(3\text{ GHz}) = 4.9$ nT²/Hz (or $\sqrt{\langle B^2 \rangle \cdot S(3\text{ GHz})} = 2.2$ nT/√Hz) can be calculated at the position closest to the write pole.

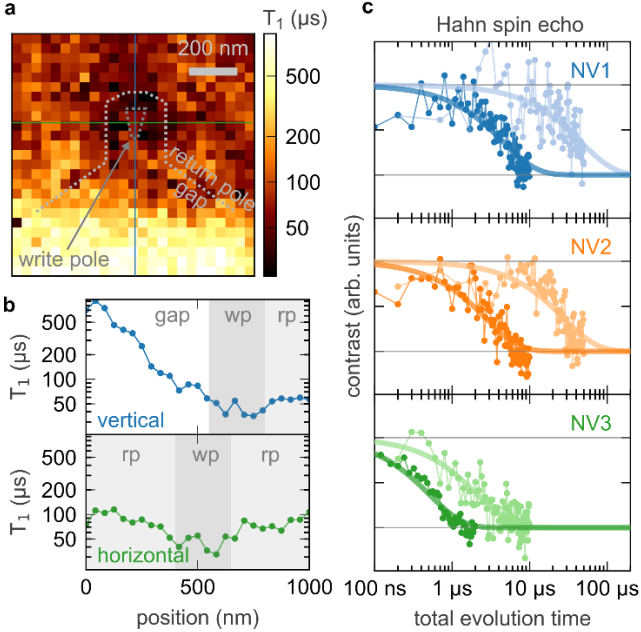

**Figure 2.** Passive effects. **(a)** $T_1$ relaxometry lateral scan of the writer. The NV's spin relaxation is measured in respect to the relative position between writerp and NV (objective fixed on NV, HDD scanned, writer off) to determine $T_1$. The estimated position of the write pole is depicted by the blue arrow. **(b)** Line scans along the lines (blue and green) in (a). The gray areas show the estimated position of the write (wp) and return pole (rp). **(c)** Comparison of the transversal relaxation time $T_2$ of three NVs (indicated as green, orange and blue) with (saturated) and without (pale) write head on top. All lines were measured with a similar overall measurement time. Therefore, lines with longer echo times appear noisier compared to lines with shorter ones.

Figure 2c shows $T_2$ measurements for three different NVs with (saturated) and without (pale) write head. We often observe transverse relaxation times close to the writer are reduced to an order of a few hundred nanoseconds (NV3, green: 510 ns), compared to a few µs without writer. In some circumstances, however, we do find relaxation times on the order of a few µs (NV1, blue: 4.6 µs and NV2, orange: 2.7 µs) with the engaged writer. When a similar analysis as for $T_1$ is applied, we can determine the magnetic noise for the second frequency domain. The filter function of a spin echo measurement depends on the total evolution time. For 10-100 µs the dominant sensitivity range is in the sub MHz regime[25]. In this range we find noise ranging from $\langle B^2 \rangle \cdot S(1\text{ MHz}) = 13$ nT²/Hz (3.6 nT/√Hz) up to 96 nT²/Hz (9.8 nT/√Hz) (NV 1 and 3 respectively).

We find that the writer introduces large magnetic noise to NV spins in its vicinity. This may impede interferometric measurements, like nuclear spin detection, or even disrupt potentially coupled networks of spins. For instance, the maximum distance at which NV1 could coherently couple to a neighboring NV would be reduced from 17 nm to 8 nm in the presence of the writer[8]. Nevertheless, while we find applications based on the spins' phase evolution to be challenging they are in principle possible.

In a similar scenario where two parallel NV are separated by 10 nm, a gradient magnetic field on the order of 100 µT/nm is required to separate the individual resonance lines by $\Delta \nu = \Delta r \, \gamma/2\pi \cdot dB/dr \approx 30$ MHz which allows for fast, high-fidelity and selective manipulations. The writer can easily produce such gradients, however, only in conjunction with strong, arbitrarily oriented fields that generally prevent ODMR. Hence, before selective microwave pulses can be applied, we must confirm that the gradient field can be switched off for initialization and read-out and that spin manipulations work in arbitrary fields. The large bandwidth of the writer allows to switch the magnetic field on fast timescales. While this is generally desired, fast magnetic field ramps can induce additional dynamics on the NV spin. One particular behavior is the non-adiabatic fast passage. If the field is misaligned from the NV symmetry axis and the Zeeman interaction exceeds the ZFS, i.e. for field strength beyond 100 mT, the eigenstates change substantially and the corresponding energies are swept over a *level anti-crossing* (LAC). In this case the spin state gradually follows the change of the eigenbasis in an adiabatic process (if the sweep is *slow* compared to the energy gap $\Delta$ at the LAC) or to a new state in a non-adiabatic process (if the sweep is *fast*). The transition probability $P$ is well described by the Landau-Zener formula[26]

$$P = e^{-\frac{\pi^2 \Delta^2}{\nu}}, \quad (3)$$

where $\nu$ is the rate of change of the eigenenergies during the sweep, also known as the Landau-Zener velocity. In the case of the NV the gap energy is given by the radial field strength $\Delta = \sqrt{2} \cdot \gamma/2\pi \cdot B_\perp$ at the avoided crossing and can be as large as the ZFS, i.e. 2.87 GHz. The Landau-Zener velocity $\nu = \gamma/2\pi \cdot \partial B_z/\partial t$ depends on the speed of the magnetic field sweep. The writer can in principle sweep the eigenenergies with velocities of GHz/ns and consequently realistically drive non-adiabatic fast passages, where $\nu \gg \Delta^2$.

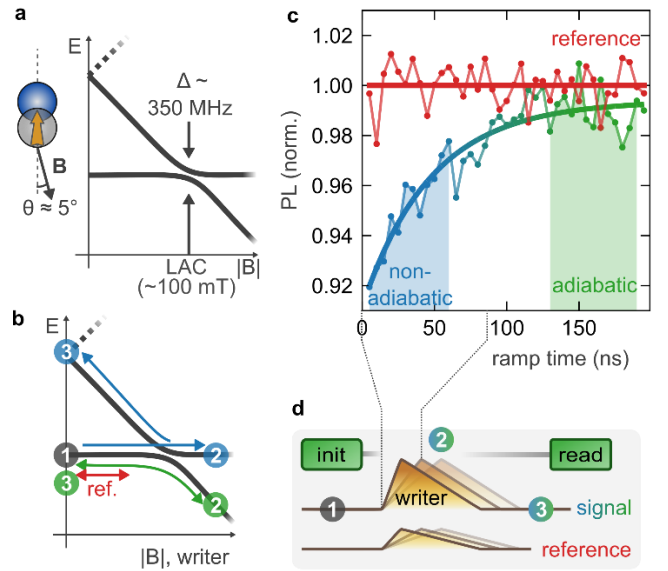

**Figure 3.** Non-adiabatic fast passage. **(a)** Level scheme for slightly misaligned magnetic fields. At about 100 mT the two lower electron spin states form a *level anti-crossing* (LAC). **(b)** A ramped magnetic field greater than 100 mT can lead to an adiabatic or non-adiabatic evolution depending on the ramp time. **(c)** For ramp times longer than 100 ns (green) the system evolves adiabatically, whereas values shorter than 50 ns (blue) lead to a non-adiabatic evolution. **(d)** Pulse scheme for the measurement in (c). For reference the red curve corresponds to a reduced ramp amplitude that prevents passing the LAC.

For an experiment we move the writer close to a NV so that its write field has a small angle $\theta < 5°$ to the NV axis and a field strength of $|B| \approx 100$ mT (see Figure 3a). This setting is close to the LAC and introduces a mixing of the $m_S = 0$ and $m_S = -1$ states. The gap energy $\Delta$ in this case is smaller than 350 MHz. We then apply a pulse sequence (see Figure 3b and d) where we polarize the NV spin into $m_S = 0$ while the writer is off. Subsequently, we ramp the current beyond the avoided crossing on a varying timescale between 5 and 200 ns. Finally, we slowly ramp it back down over a period of 1 µs and read out the spin state. Assuming the falling ramp to be slow enough to change the spin state adiabatically, we expect fast passages during the rising edge to manifest as a reduction in fluorescence. Figure 3c shows the result of the measurement. As expected we find non-adiabatic fast-passages for short rising edges and an exponential behavior towards an adiabatic sweep. As a reference we also recorded sweeps at lower current amplitudes that do not drive the system beyond the LAC. In this case we do not observe any population transfer, which is consistent with the theory.

Assuming a linear increase of the magnetic field strength during the ramp, i.e. $B_z(t) = B_{max} \cdot t/t_{ramp}$, the Landau-Zener formula

$$P = e^{-\frac{\pi^2 \Delta^2}{v}} = e^{-\frac{\pi^2}{\gamma/2\pi} \cdot \frac{\Delta^2}{B_{max}} \cdot t_{ramp}} = e^{-\frac{t_{ramp}}{\tau}} \quad (4)$$

is expressed in terms of the maximum field strength $B_{max}$ and the ramp time $t_{ramp}$. This formula is used to fit the measurement data in Figure 3c by introducing and adjusting the fit parameter $\tau = \frac{\gamma/2\pi}{\pi^2} \cdot \frac{B_{max}}{\Delta^2} = 48$ ns which depends on both $B_{max}$ and $\Delta$.

The exact value of the energy splitting $\Delta$ during the sweep is unknown, as we can only characterize static fields for corresponding DC currents. The dynamic field however may not necessarily have the same orientation as the settled system. The same argument holds for the maximum magnetic field strength $B_{max}$. As the magnetic field has to exceed at least the LAC at the end of the sweep, a realistic lower bound for $B_{max}$ can be estimated. By assuming a $B_{max} = 100$ mT and taking the fit parameter $\tau = 48$ ns into account also a lower bound for the energy splitting of $\Delta = 77$ MHz can be calculated. The corresponding maximum magnetic field rate $\dot{B}_z = \partial B_z / \partial t$ for $t_{ramp} = 5$ ns is $\dot{B}_z = 20$ mT/ns. For a realistic value the maximum magnetic field strength of about $B_{max} = 200$ mT measured with an NV for hard disk recording heads[10] can be considered. This leads to a splitting of $\Delta = 109$ MHz and a magnetic field rate of $\dot{B}_z = 40$ mT/ns ($t_{ramp} = 5$ ns).

In principle Landau-Zener transitions are coherent processes. By finding the proper ramp times $t_{ramp}$, gates similar to $\pi$- or, in particular, $\pi/2$-pulses are conceivable. In our experiments we were, however, not able to observe coherent state evolutions. Small changes in the magnetic field settings immediately translate into vastly different evolution speeds. We suspect that we could not drive the magnetic field accurately enough to a consistent setting and hence blurred out any coherent evolution under a large inhomogeneous broadening.

With the knowledge of the spin's behavior during magnetic field ramps, we can now construct a scheme to manipulate and retrieve the spin state in arbitrary fields and characterize the gradient field, i.e. search for appropriate resonances for selective addressing. As long as the magnetic field is ramped slowly enough for the spin to transition adiabatically from one quantization basis to the other, i.e. $\gamma/2\pi \cdot \dot{B} \ll \Delta^2$, the spin states populations are mapped unambiguously from basis to basis. As shown in the schematic in Figure 4a, this allows to probe for spin resonances in the radial field and retrieve the spin state optically. In an experiment we bring the write pole close to an NV defect. We apply a pulse scheme as shown in Figure 4c, where a laser pulse initializes the spin before the current on the writer is ramped up to 10-20 mA within 100 ns. The magnetic field is kept static for the duration of a probing microwave pulse before it is ramped down again for the optical readout. Figure 4b shows a spectrum recorded with this method. From the resonances we derive a magnetic field of $|B| = 75.4$ mT and a polar angle $\theta = 52°$ during the write pulse. For larger fields the spin transitions between the highest and lowest states may become forbidden. In such a case additional microwave pulses are required to observe both transitions.

Expanding on this method, we move the writer within its accuracy by 25 nm steps to map out the magnetic field around the write pole. Figure 4d shows the magnetic field strength and the axial and radial components that we derive from the measurement. As expected we see the magnetic field to vary strongly over the range of a few hundred nanometers ranging from 15 to 28 mT. Furthermore, we find magnetic field gradients on the order of 100 µT/nm, which is comparable to the gradient of magnetized atomic force microscope tips or nanostructures for magnetic resonance force microscopes[27,28].

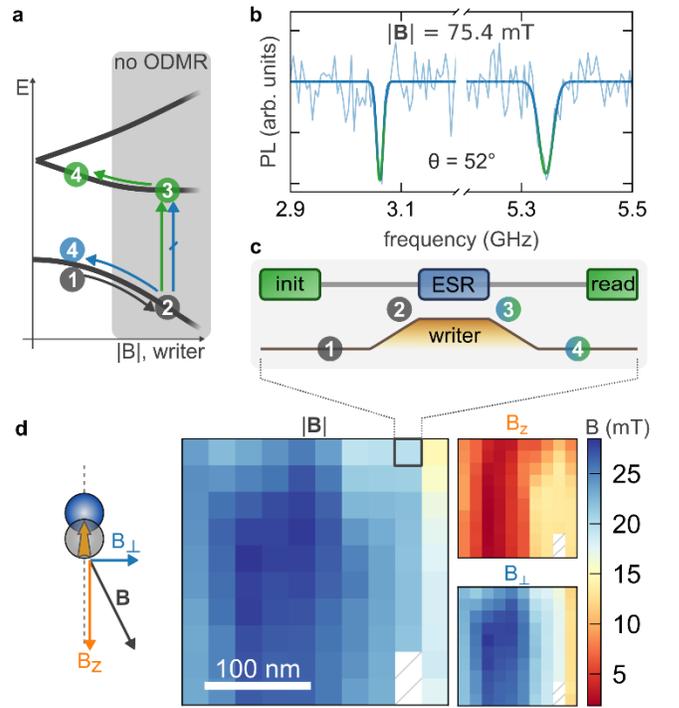

**Figure 4.** Pulsed magnetic write fields. **(a)** Pulsed field ODMR. Due to the large bandwidth the write field can be applied only for a $\pi$-pulse while it is off for initialization and read-out. That way spin resonances can be probed at magnetic field strengths much greater than for DC magnetic fields. **(b)** Spectrum with two resonances corresponding to a magnetic field strength of 75.4 mT with a misalignment of 52°. The coloring indicates the path corresponding to the scheme in (a) in case the microwave is resonant (green) or off-resonant (blue). **(c)** Pulse scheme for the measurements in (b) and (d). **(d)** Magnetic field lateral scan around the write pole obtained by pulsed field ODMR. The magnetic field strengths range from 15 mT to 28 mT leading to magnetic field gradients on the order of 100 µT/nm. The spin transitions can be observed and controlled in arbitrarly oriented magnetic fields.

Finally, this new ODMR method allows using the gradient field to address individual parallel NVs on the nanoscale[7,27,29]. Here we move the writer over a detected NV fluorescence spot that contains such a pair of NVs (see Figure 5a and b) and apply the same pulse sequence as before (see Figure 4) with moderate current amplitudes. At low currents we first find two resonance lines corresponding to the two spin transitions of both defects (Figure 5c). However, when the current is increased the lines split into two pairs of lines corresponding to the individual transitions of each defect. In this case each NV spin state can be manipulated individually with a fidelity of 99% (calculated by the spectral overlap) and 10 MHz Rabi frequencies for the left transition pair in Figure 5c. By reducing the power broadening or applying higher write currents (corresponding to higher magnetic field gradients) the minimal distance of spectrally addressable parallel NV pairs can be even further reduced.

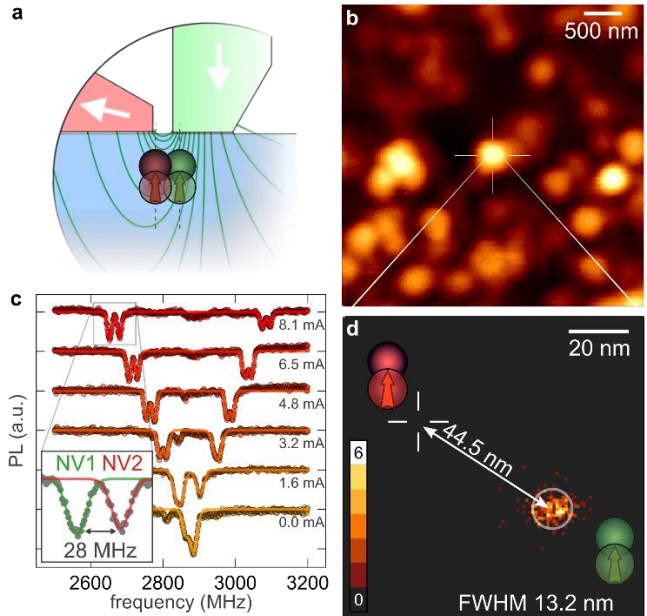

**Figure 5.** Selective addressing of a pair of NVs using tunable magnetic field gradients. **(a)** Schematic picture of two parallel NVs in the gradient field of the writer. **(b)** A confocal scan shows NVs as diffraction limited spots. The central spot contains a parallel pair. **(c)** ODMR spectra of the NV pair. At low currents the spectral lines overlap. With increasing current first two resonance lines appear corresponding to the two possible transitions of both NVs. If the current is further increased the lines split up into two pairs of lines corresponding to the individual transitions of each defect. The fidelity, calculated by the spectral overlap, is 99% for the left transition pair. **(d)** Optically reconstructed relative (lateral) position of the parallel NVs in the centered confocal spot. The figure shows a histogram (1 nm bin width) of 286 individual DESM scans. The crosshair marks the origin point, i.e. the position of NV1. The relative distance to NV2 of 44.5 nm is determined with a resolution (FWHM) of 13.2 nm.

The distance between the addressed NVs is an important measure for the applicability on spin networks. We determine their relative position with an optical reconstruction method, called *deterministic emitter switch microscopy* (DESM)[30] and at the same time demonstrate the usefulness of our addressing method. This super-resolution method requires selective microwave pulses to switch individual defects to its dark spin state so that its image can be separated from a reference image where all emitters are bright. Figure 5d shows a histogram of 286 relative positions of the pair extracted from DESM scans. We find a relative distance of 44.5 nm with a resolution (FWHM) of 13.2 nm. That implies that spectral addressing is possible on length scales close to typical dipolar coupling ranges. As we did not have to use the full strength of the gradient the approach would likewise work on defects as close as 10 nm.

Our experiments show that HDD writers are a valuable tool for the generation of control fields in spin experiments. We are able to use various new degrees of freedom which are enabled by their remarkable features. The GHz bandwidth allows to rapidly pulse control fields, modulate the spin dynamics in run time and drive non-adiabatic fast passages. Pulsed magnetic fields allow measurements in strong off-axial fields that would otherwise not yield sufficient fluorescence contrast. This enables the use of the magnetic field gradient to address individual NV spins on the nanoscale. The method could easily be expanded to address larger clusters of NVs. However, the introduction of magnetic noise from the writer puts additional limitations on the interaction range of a coupled network. Yet, as the writer can produce gradients on the order of mT/nm, spins within a few nanometer distance can be addressed. Hence, HDD writers are suitable for applications controlling large spin arrays as for example in proposed diamond-based quantum processor[2,7,8]. Ultimately nano-fabricated writers can serve as a template for micro- and nanostructures specifically tailored to the requirements of single spin research and devices.

## AUTHOR INFORMATION

### Corresponding Author
*E-mail: i.jakobi@physik.uni-stuttgart.de.

### Author Contributions
S.B., I.J., J.M., I.G., P.N. and J.W. conceived the experiments. S.B. and I.J. performed the experiments. S.B. and I.J. analyzed the data. S.B., I.J. and J.W. wrote the manuscript.

### Notes
The authors declare no competing financial interests.

## ACKNOWLEDGMENT

We acknowledge financial support by the German Science Foundation (SPP1601 and FOR1493), the EU (ERC grant SMeL), the Volkswagen Foundation, the Humboldt Foundation, the Baden Wuerttemberg Foundation and the MPG.
Furthermore, we thank Fadi El Hallak of Seagate Technology for providing hard disk head samples and technical assistance and Robert McMichael of NIST CNST for fruitful discussions.